\documentstyle[multicol,prl,aps,epsfig]{revtex}

\begin{document}
\draft

\title{Observation of a Single-Spin Azimuthal Asymmetry in 
Semi-Inclusive Pion Electro-Production }
\author{
A.~Airapetian$^{32}$, 
N.~Akopov$^{32}$,
M.~Amarian$^{24,27,32}$, 
E.C.~Aschenauer$^{13,14,6}$, 
H.~Avakian$^{10}$, 
R.~Avakian$^{32}$, 
A.~Avetissian$^{32}$, 
B.~Bains$^{15}$,
C.~Baumgarten$^{22}$,
M.~Beckmann$^{12}$, 
S.~Belostotski$^{25}$, 
J.E.~Belz$^{28,29}$,
Th.~Benisch$^{8}$, 
S.~Bernreuther$^{8}$, 
N.~Bianchi$^{10}$,
J.~Blouw$^{24}$, 
H.~B\"ottcher$^{6}$, 
A.~Borissov$^{6,14}$, 
M.~Bouwhuis$^{15}$,
J.~Brack$^{4}$,
S.~Brauksiepe$^{12}$,
B.~Braun$^{22,8}$, 
B.~Bray$^{3}$,
St.~Brons$^{6}$,
W.~Br\"uckner$^{14}$, 
A.~Br\"ull$^{14}$,
E.E.W.~Bruins$^{19}$,
H.J.~Bulten$^{18,24,31}$,
G.P.~Capitani$^{10}$, 
P.~Carter$^{3}$,
P.~Chumney$^{23}$,
E.~Cisbani$^{27}$, 
G.R.~Court$^{17}$, 
P.~F.~Dalpiaz$^{9}$, 
E.~De Sanctis$^{10}$, 
D.~De Schepper$^{19,2}$, 
E.~Devitsin$^{21}$, 
P.K.A.~de Witt Huberts$^{24}$, 
P.~Di~Nezza$^{10}$,
M.~D\"uren$^{8}$, 
A.~Dvoredsky$^{3}$, 
G.~Elbakian$^{32}$,
J.~Ely$^{4}$,
A.~Fantoni$^{10}$, 
A.~Fechtchenko$^{7}$,
M.~Ferstl$^{8}$,
K.~Fiedler$^{8}$, 
B.W.~Filippone$^{3}$, 
H.~Fischer$^{12}$, 
B.~Fox$^{4}$,
J.~Franz$^{12}$, 
S.~Frullani$^{27}$, 
M.-A.~Funk$^{5}$, 
Y.~G\"arber$^{6}$, 
H.~Gao$^{2,15,19}$,
F.~Garibaldi$^{27}$, 
G.~Gavrilov$^{25}$, 
P.~Geiger$^{14}$, 
V.~Gharibyan$^{32}$,
A.~Golendukhin$^{8,22,32}$, 
G.~Graw$^{22}$, 
O.~Grebeniouk$^{25}$, 
P.W.~Green$^{1,29}$, 
L.G.~Greeniaus$^{1,29}$, 
C.~Grosshauser$^{8}$,
M.~Guidal$^{24}$,
A.~Gute$^{8}$,
V.~Gyurjyan$^{10}$,
J.P.~Haas$^{23}$,
W.~Haeberli$^{18}$, 
J.-O.~Hansen$^{2}$,
M.~Hartig$^{29}$, 
D.~Hasch$^{10}$,
O.~H\"ausser\cite{author_note1}$^{28,29}$,
F.H.~Heinsius$^{12}$,
R.~Henderson$^{29}$,
M.~Henoch$^{8}$, 
R.~Hertenberger$^{22}$, 
Y.~Holler$^{5}$, 
R.J.~Holt$^{15}$, 
W.~Hoprich$^{14}$,
H.~Ihssen$^{5,24}$, 
M.~Iodice$^{27}$, 
A.~Izotov$^{25}$
% ...
\centerline {\it (The HERMES Collaboration)}
}

\address{
% ...
$^5$DESY, Deutsches Elektronen Synchrotron, 22603 Hamburg, Germany\\
% ...
$^8$Physikalisches Institut, Universit\"at Erlangen-N\"urnberg, 91058 Erlangen, Germany\\
% ...
$^{12}$Fakult\"at f\"ur Physik, Universit\"at Freiburg, 79104 Freiburg, Germany\\
% ...
$^{14}$Max-Planck-Institut f\"ur Kernphysik, 69029 Heidelberg, Germany\\ 
$^{15}$Department of Physics, University of Illinois, Urbana, Illinois 61801, USA\\
% ...
$^{22}$Sektion Physik, Universit\"at M\"unchen, 85748 Garching, Germany\\
% ...
$^{24}$Nationaal Instituut voor Kernfysica en Hoge-Energiefysica (NIKHEF), 1009 DB Amsterdam, The Netherlands\\
$^{25}$Petersburg Nuclear Physics Institute, St. Petersburg, 188350 Russia\\
% ...
$^{27}$Istituto Nazionale di Fisica Nucleare, Sezione Sanit\`a and Physics Laboratory, Istituto Superiore di Sanit\`a, 00161 Roma, Italy\\
$^{28}$Department of Physics, Simon Fraser University, Burnaby, British Columbia V5A 1S6, Canada\\ 
$^{29}$TRIUMF, Vancouver, British Columbia V6T 2A3, Canada\\
% ...
}
\date{\today}
\maketitle

\begin{abstract}

Single-spin asymmetries for
semi-inclusive pion production in 
deep-inelastic scattering have been measured for the first time.
A significant target-spin asymmetry of the distribution in the azimuthal 
angle $\phi$ of the pion relative
to the lepton scattering plane
was observed
for $\pi^+$  electro-production on a longitudinally polarized hydrogen target.
The corresponding 
analyzing power in the $\sin \phi$ moment of the cross section
is  $0.022 \pm 0.005 \pm 0.003$.
This result can be interpreted as the effect of terms in the cross section
involving chiral-odd spin distribution functions in combination 
with a time-reversal-odd fragmentation function that is sensitive
to the transverse polarization of the fragmenting quark. 

\end{abstract}


\begin{multicols}{2}[]

Polarized deep-inelastic lepton scattering has 
been the primary experimental basis for our
present understanding of the spin structure of the nucleon. 
Inclusive and semi-inclusive measurements 
with both beam and target polarized 
have been used to provide
precise information on quark helicity-distribution functions.
These quantities represent the
distribution of quark spin in a longitudinally polarized nucleon,
in a suitably Lorentz-boosted kinematic frame.
Additional spin-distribution functions 
have been identified, but remain unmeasured.
One of these is called transversity and corresponds to the 
distribution of transverse quark spin in a nucleon
polarized transverse to its (infinite) momentum~\cite{JAFJ}.
This and related distribution functions
are predicted to be measurable via single-spin asymmetries, where only
the beam {\it or} target are polarized, in certain 
lepton and hadron scattering experiments~\cite{COL,AK,TM,JAF1,ANS1}.

In simple models based on hadrons consisting of non-interacting
collinear partons
(quarks and gluons), single-spin asymmetries are expected to vanish.
This follows from the conservation of parity, total
angular momentum and  helicity of the individual partons.
Correspondingly, in the language of perturbative QCD, 
single-spin asymmetries vanish at the ``twist-2" level, i.e. when
multi-parton correlations and parton transverse momenta internal to hadrons 
are ignored.
However, single-spin asymmetries have been observed
in a few hadron-hadron
scattering experiments~\cite{ADA}. In these measurements, a scattered hadron
was detected with a momentum transverse to the beam direction in the range
$P_\perp\simeq$ 1-2 GeV,
which is not much larger than either the scale parameter of QCD 
($\Lambda_{QCD}\sim 0.2$ GeV)
or typical parton transverse momenta of a few hundred MeV.
Therefore these asymmetries may
arise from non-collinear parton configurations or from multi-parton 
correlations (``higher twist'' effects), which are suppressed at 
large  $P_\perp$ where perturbative QCD becomes effective.

Transversity and related spin-distribution functions are as yet unmeasured 
because their unusual chiral-odd structure implies that 
they are not directly observable in inclusive lepton-nucleon scattering 
experiments~\cite{JAFJ}.
However, it has been suggested that the needed sensitivity can be provided 
by semi-inclusive production of pions with modest $P_\perp$~\cite{COL}.
An observable  single-spin dependence is predicted to appear 
in the dependence of the cross section on
the angle between the  spin axis of a {\it transversely} polarized target and the
plane defined by the virtual photon
momentum and the momentum  of the pion (known as the Collins angle). 
Here the pion is produced from the struck quark in soft processes
described by a fragmentation function
having a chiral-odd structure 
like that of the spin-distribution functions of interest. This 
Collins 
fragmentation function describes how the probability for producing
a pion depends on its direction with respect to the direction of 
transverse polarization of the struck quark. 
It has also a time-reversal odd structure
resulting from the final-state interactions in
the fragmentation process, rather than from any fundamental violation of
time-reversal invariance~\cite{gasi}. 
Such T-odd fragmentation
(and distribution) functions can thus be considered as effective
parameterizations of specific complex processes.
There is preliminary evidence from Z$^0 \rightarrow 2$-jet decay
\cite{EFR} that 
the Collins fragmentation 
function has a substantial magnitude --
of order 10\% of the well-known chiral-even spin-independent one. 
If this can be confirmed, it could provide experimental sensitivity
to the transverse polarization of scattered quarks in future experiments 
designed to make the first measurements of transversity.

In the case of semi-inclusive pion production in lepton scattering 
from a {\em longitudinally} polarized nucleon, chiral-odd
quark spin-distribution functions closely related to transversity
can be manifest.
In such experiments, the Collins angle becomes
the azimuthal angle $\phi$ of the pion around the virtual photon direction,
with respect to the lepton scattering plane. 
Recent theoretical studies~\cite{AK,TM} have shown how
each chiral-odd spin-distribution function coupled with the 
Collins fragmentation function 
gives rise to a specific single-spin dependent moment of the 
pion yield distribution in $\phi$.

\begin{figure}[H]
  \begin{center}
   \epsfig{file=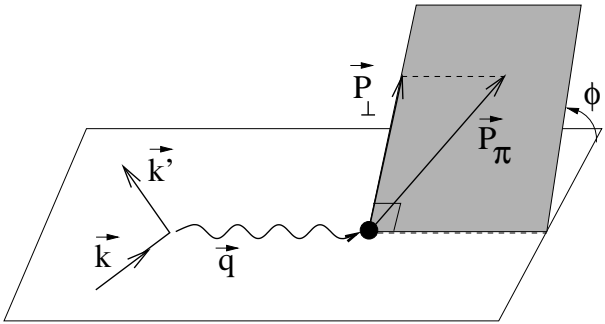,width=8cm}
   \end{center}
\begin{minipage}[r]{\linewidth}
   \caption{Kinematic planes for pion production in semi-inclusive deep-inelastic scattering.}
 \label{fig:kinemplane}
\end{minipage}
 \end{figure}

The kinematics of the process are illustrated in Fig.~\ref{fig:kinemplane}.
The relevant variables are the 4-momentum transfer squared 
$-Q^2=q^2=(k-k^{'})^2$,
the energy transfer $\nu=E-E^{'}$, 
the virtual photon fractional energy $y=\nu/E$, 
the invariant mass of the photon-proton system $W=\sqrt{2M\nu+M^2-Q^2}$,
the Bjorken variable $x=Q^2/2M\nu$, and
the pion fractional energy $z=E_{\pi}/\nu$.
Here $k$ and $k^{'}$ are the 4-momenta and $E$ and $E^{'}$
the laboratory energies of the incoming and outgoing leptons, respectively.
$E_{\pi}$ is the pion laboratory energy and $M$ the proton mass.
The transverse momentum ($P_{\perp}$) of the pion
is defined with respect to the
virtual photon direction in the initial photon-proton center-of-mass
frame.

This Letter reports the first observation of a 
single-spin azimuthal asymmetry for semi-inclusive pion production 
in deep-inelastic scattering. 
The data were recorded during the 1996 and 1997 running periods of the HERMES
experiment using both unpolarized and longitudinally nuclear-polarized  
hydrogen internal gas targets \cite{STOCK} in the 27.6 GeV HERA 
polarized positron storage ring at DESY. 
Longitudinal beam polarization is obtained by using spin rotators\cite{BAR}
located upstream and downstream of the HERMES experiment.
The scattered positrons and associated pions 
are detected by the HERMES spectrometer \cite{HD} in the
polar angle range 0.04 rad $< \theta<$ 0.22 rad. Positron and hadron
identification is based on information from four
detectors: a threshold gas \v{C}erenkov counter, a transition-radiation detector,
 a preshower scintillator detector and  a
lead-glass electromagnetic calorimeter.
The particle identification provides an
average positron identification efficiency of 99\% with a hadron
contamination that is
less than 1\%.

The kinematic requirements on the scattered positron used in this analysis are
1 GeV$^2 <Q^2<$ 15 GeV$^2$, $W>$ 2 GeV,
0.023 $<x<$ 0.4 and
$y <$ 0.85.
Pions were identified in the energy range  $4.5$ GeV $<E_{\pi}<13.5$  GeV.
Acceptance effects were minimized and exclusive production was suppressed
by imposing the requirement 0.2 $<z<$ 0.7.
The limit $P_{\perp}>50$ MeV was applied
to the pions to allow an accurate measurement 
of the angle $\phi$.

Measurements were performed with all combinations of
beam and target helicities, 
giving the possibility of measuring single- and 
double-spin terms in the cross section.
The average hydrogen target polarization in the 1996 and 1997
HERMES running periods was 0.86 with a fractional uncertainty of 5\%.
The average beam polarization for the analyzed data was 0.55
with a fractional uncertainty of 3.4\%.

The various contributions to the
$\phi$ dependent spin asymmetry are isolated by
extracting moments of the cross section weighted by 
corresponding $ \phi $ dependent functions.
The analyzing powers for beam (target) longitudinal polarization
are evaluated as 
\begin{equation}
\label{sinfi}
A^W_{LU(UL)}=
\frac{\frac{L^\uparrow}{L^\uparrow_P}\sum_{i=1}^{N^\uparrow} W(\phi_i^\uparrow) \ -\
      \frac{L^\downarrow}{L^\downarrow_P}\sum_{i=1}^{N^\downarrow} W(\phi_i^\downarrow)} 
     {\frac{1}{2} [N^\uparrow\ +\ N^\downarrow]}, 
\end{equation}
where the $\uparrow /\downarrow$ denotes positive/negative
helicity of the beam (target). 
$N^{\uparrow /\downarrow}$ is the number of selected events involving a
detected pion for each beam (target) spin state corresponding to
the dead-time corrected luminosities $L^{\uparrow /\downarrow}$
and $L^{\uparrow /\downarrow}_P$, the latter being averaged with
the magnitude of the beam (target) polarization.
All of these quantities are effectively averaged over the two 
target (beam) helicity states to arrive at single-spin asymmetries.
The weighting functions $W(\phi)=\sin \phi$ and $W(\phi)=\sin2\phi$ 
are expected to provide sensitivity to the Collins fragmentation function
discussed above, in combination with 
different spin distribution functions~\cite{AK,TM}.
Analyzing powers were extracted 
by integrating over the spectrometer acceptance in the kinematic variables
$y$ and $z$.
Corrections were applied for the effects of the spectrometer acceptance,
based on a Monte Carlo simulation.

The values of  ${\it A}^{\sin \phi}_{UL}$, ${\it A}^{\sin 2\phi}_{UL}$
and   ${\it A}^{\sin \phi}_{LU}$ extracted from the data according to 
Eq.~(\ref{sinfi}) and averaged over $x$ and $P_{\perp}$ are given 
 in Table~\ref{Tab2}. For both  ${\pi^+}$ and ${\pi^-}$ the beam-related analyzing powers
 ${\it A}^{\sin \phi}_{LU}$ are consistent with zero. 
This is in agreement with the small contributions to  
${\it A}^{\sin \phi}_{LU}$ predicted to arise from 
higher-twist and $\it O$$(\alpha_S^2$)  
QCD effects \cite{LEV,F5}.
The target-related term ${\it A}^{\sin 2\phi}_{UL}$
is also consistent with zero within errors, both for
 ${\pi^+}$ and ${\pi^-}$.

The other target-related analyzing power ${\it A}^{\sin \phi}_{UL}$
is  consistent with zero for ${\pi^-}$, while it is
significantly different from zero for ${\pi^+}$. 
The appearance of such an asymmetry
suggests the influence of the Collins T-odd fragmentation function,
in combination with one or more chiral-odd spin-distribution functions.
The large difference between the $\pi^+$ and $\pi^-$ asymmetries
can be understood (in the QPM) only if there is a large difference between
favored and disfavored chiral-odd fragmentation functions --
{\it i.e.} if there is a strong enhancement when the struck quark flavor 
is present in the hadron.

\begin{figure}[H]
  \begin{center}
   \epsfig{file=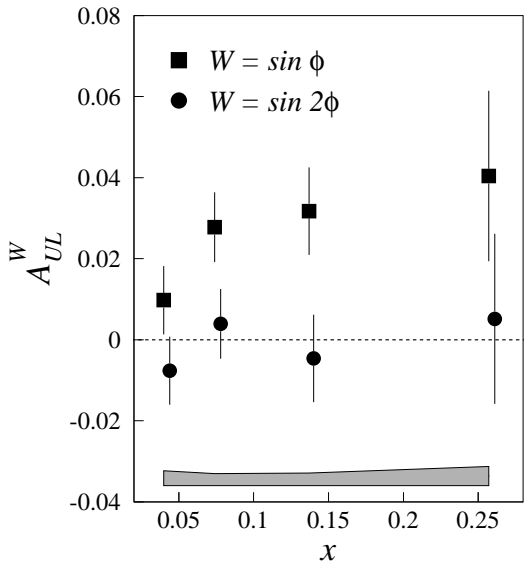,width=7cm}
   \end{center}
\begin{minipage}[r]{\linewidth}
   \caption{
Target-spin analyzing powers for $\pi^+$: 
${\it A}^{\sin\phi}_{UL}$ (squares) and  
${\it A}^{\sin 2\phi}_{UL}$ (circles)
as a function of Bjorken $x$.  
 Error bars show the
statistical uncertainty and the band 
represents the systematic uncertainties for ${\it A}^{\sin\phi}_{UL}$.  
As shown in Table~\ref{Tabxdis}, $\langle Q^2 \rangle$ varies with $x$.
  }
 \label{fig:azimpolxxpipp7100}
\end{minipage}
 \end{figure} 

In Table~\ref{Tabxdis} the ${\it A}^{\sin\phi}_{UL}$ and ${\it A}^{\sin 2\phi}_{UL}$ analyzing powers are given for $\pi^+$ and $\pi^-$ at  
the measured $\langle x \rangle$ and  $\langle Q^2 \rangle$ values. In addition,
in Fig.~\ref{fig:azimpolxxpipp7100}, the  
${\it A}^{\sin\phi}_{UL}$ and ${\it A}^{\sin 2\phi}_{UL}$  
values obtained for $\pi^+$  are  shown as a function of $x$,
after averaging over $P_{\perp}$.
At higher energies, the analyzing power for the 
$\sin \phi$ moment that is sub-leading order in $1/Q$
is expected to be suppressed by the factor of $P_{\perp}/Q$ \cite{AK,TM} 
with respect to  the leading-order $\sin 2\phi$ moment. 
In the HERMES kinematics, which covers a range of relatively low $Q^2$ 
and moderate $P_{\perp}$, 
the ratio of ${\it A}^{\sin2\phi}_{UL}$ to
${\it A}^{\sin\phi}_{UL}$ analyzing
powers is predicted to be small
in the valence region \cite{MBOG}. This is in agreement with a
simple estimate of that ratio in the real photon limit \cite{TER}. 
The present data are consistent with these theoretical expectations,
neglecting the contribution to the $\sin \phi$ moment from transversity
itself arising from the small component of the target spin transverse to the
virtual photon direction.
Also, the apparent increase of ${\it A}^{\sin\phi}_{UL}$ with increasing $x$
suggests that the sea contribution does not dominate the 
effect,  in agreement with existing interpretations of
single-spin asymmetries as being associated with valence quark 
contributions \cite{ARTRU,ANSBOG}. 

\begin{figure}
  \begin{center}
   \epsfig{file=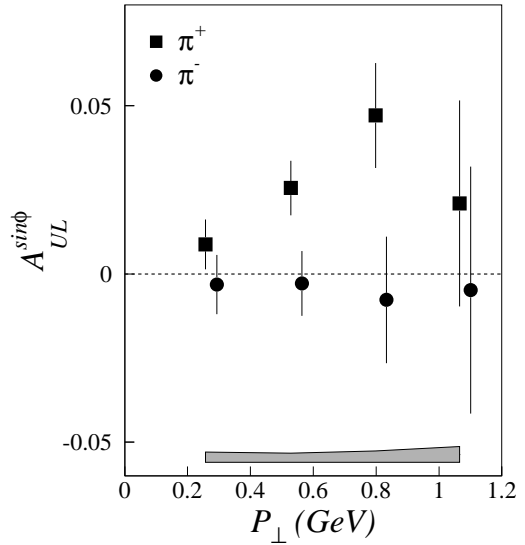,width=7cm}
   \end{center}
\begin{minipage}[r]{\linewidth}
   \caption{ 
Target-spin analyzing powers in the $\sin \phi$ moment
as a function of transverse momentum,
for $\pi^+$ (squares) and  $\pi^-$ (circles).
 Error bars show the
statistical uncertainties and the band 
represents the systematic uncertainties.
}
 \label{fig:azimpolptxpipp7100}
\end{minipage}
 \end{figure}

In Fig.~\ref{fig:azimpolptxpipp7100},
${\it A}^{\sin\phi}_{UL}$ averaged over $x$
is plotted for $\pi^+$ and  $\pi^-$ as a function  of
transverse momentum. The mean  $\langle Q \rangle$ is about 1.55 GeV 
for all bins.
There is an indication that ${\it A}^{\sin\phi}_{UL}$ for $\pi^+$
increases as $P_\perp$ increases up to $\sim$0.8 GeV. This behavior can be 
related to the dominant role of the 
intrinsic quark transverse momentum when $P_\perp$ remains below
a typical hadronic mass ($\sim $ 1 GeV).
On this basis, the use of Gaussian 
transverse momentum parameterizations for distribution
and fragmentation functions 
results in a  behavior of ${\it A}^{\sin\phi}_{UL}$ 
that is proportional to $P_\perp$, at least for the moderate
range of $P_\perp$ \cite{AK,TM,OABK}.     

 The main contributions to the systematic uncertainties are those from
 the target and beam polarizations, from smearing due to detector resolution and from
 a false spin asymmetry induced by the spectrometer acceptance.
Uncertainties in the acceptance corrections based on Monte Carlo calculations 
dominate the systematic uncertainties 
at small $x$ and decrease
with increasing $x$.
At the average values of $y$ of about 0.5, radiative
effects are expected to be small and independent of the
pion charge; 
these effects on the unpolarized cross section were
evaluated and were indeed found to be negligible~\cite{AKU}.

In summary, single-spin azimuthal asymmetries of pions produced
in deep-inelastic scattering of polarized positrons
from a longitudinally polarized hydrogen target have been measured. 
The analyzing power involving the $\sin\phi$ moment of the cross section
is found to be significant for
$\pi^+$-production with unpolarized (spin-averaged) positrons 
on a longitudinally polarized hydrogen target,
while for $\pi^-$ it is found to be
consistent with zero. In addition, the analyzing powers involving the
$\sin 2\phi$ moments of both
$\pi^+$ and $\pi^-$ are consistent with zero.
The $\sin\phi$ target-related analyzing power for $\pi^+$, averaged over the full 
acceptance, is found to be 
$0.022\pm0.005\pm0.003$, and there are indications that 
this analyzing power increases with
increasing $x$, and also with $P_\perp$ up to $\sim$0.8 GeV.
The appearance of this single-spin 
asymmetry can be interpreted as an effect of chiral-odd spin
distribution functions coupled with a time-reversal-odd fragmentation
function. 
This fragmentation function offers a means to measure transversity 
in future experiments using a {\em transversely} polarized target.

We thank M.~Anselmino, J.~Collins, A.M.~Kotzinian and P.J.~Mulders 
for many interesting discussions.

We gratefully acknowledge the DESY management for its support, the 
staffs at DESY and the collaborating institutions for their significant 
effort, and our funding agencies for financial support. 
This work was supported by the Fund for Scientific Research-Flanders (FWO) 
of Belgium; the Natural Sciences and Engineering Research Council of Canada;
the INTAS, HCM and TMR contributions from the European Community;
the German Bundesministerium f\"{u}r Bildung, Wissenschaft, Forschung 
und Technologie (BMBF), the Deutscher Akademischer Austauschdienst (DAAD); 
the Italian Istituto Nazionale di Fisica Nucleare (INFN);
Monbusho, JSPS, and Toray Science Foundation of Japan;
the Dutch Stichting voor Fundamenteel Onderzoek der Materie (FOM);
the UK Particle Physics and Astronomy Research Council; and
the US Department of Energy and National Science Foundation.

\begin{table}
\begin{minipage}[r]{\linewidth}
\begin{tabular}{|c|r|r|}
 & \multicolumn{1}{c|}{$\pi^+$}  & \multicolumn{1}{c|}{$\pi^-$} \\
\hline
${\it A}^{\sin\phi}_{UL} $  & 0.022$\pm$0.005$\pm$0.003 & -0.002$\pm$0.006$\pm$0.004 \\
${\it A}^{\sin2\phi}_{UL}$  & -0.002$\pm$0.005$\pm$0.010 & -0.005$\pm$0.006$\pm$0.005 \\
${\it A}^{\sin\phi}_{LU} $  & -0.005$\pm$0.008$\pm$0.004 &-0.007$\pm$0.010$\pm$0.004 \\
\end{tabular}
\caption{Target- and beam-related analyzing powers,
averaged over $x$ and $P_{\perp}$,
for the azimuthal $\sin \phi$ and $\sin 2\phi$
moments of the pion production cross section in deep-inelastic scattering.} 
\label{Tab2}
\end{minipage}
\end{table}
\end{multicols}
\mediumtext
\begin{table}
\begin{tabular}{|c|c|r|r|r|r|}
 & $\langle Q^2 \rangle$ & \multicolumn{2}{c|}{$\pi^+$} &
 \multicolumn{2}{c|}{$\pi^-$} \\ \cline{3-6} \rule{0mm}{4.0mm} 
 $\langle x \rangle$  & $(GeV^2)$ &
 \multicolumn{1}{c|}{${\it A}^{\sin\phi}_{UL}$} 
 & \multicolumn{1}{c|}{${\it A}^{\sin2\phi}_{UL}$}  
 & \multicolumn{1}{c|}{${\it A}^{\sin\phi}_{UL}$} 
 & \multicolumn{1}{c|}{${\it A}^{\sin2\phi}_{UL}$}    \\
\hline
  0.040  &  1.4  & 0.010$\pm$0.008$\pm$0.004 & -0.008$\pm$0.008$\pm$0.011  
& -0.004$\pm$0.010$\pm$0.004 &  0.002$\pm$0.010$\pm$0.008  \\
  0.074  &  2.2  & 0.028$\pm$0.009$\pm$0.003 &  0.007$\pm$0.009$\pm$0.012  
& -0.004$\pm$0.010$\pm$0.003 & -0.008$\pm$0.010$\pm$0.010   \\
  0.137  &  3.7  & 0.032$\pm$0.011$\pm$0.003 & -0.005$\pm$0.011$\pm$0.009  
&  0.012$\pm$0.013$\pm$0.003 & -0.007$\pm$0.013$\pm$0.007   \\
  0.257  &  6.4  & 0.041$\pm$0.021$\pm$0.005 &  0.005$\pm$0.021$\pm$0.009  
& -0.025$\pm$0.028$\pm$0.005 & -0.028$\pm$0.028$\pm$0.008   \\
\end{tabular}
\caption{Target-related analyzing powers 
averaged over $P_{\perp}$,
for the azimuthal $\sin \phi$ and $\sin 2\phi$
moment of the $\pi^+$ and  $\pi^-$ production cross section 
in deep-inelastic scattering as a function of $x$.} 
\label{Tabxdis}
\end{table}
\end{document}